\documentclass{emulateapj}
\pdfoutput=1
\usepackage{natbib}
\usepackage{microtype}
\usepackage{graphicx}
\usepackage{textcomp}
\usepackage{gensymb}
\usepackage{color}

\newcommand{\arcsecs}{\hbox{$^{\prime\prime}$}}

\shorttitle{UPWs into the corona}
\shortauthors{Freij et al.}

\begin{document}

\title{The detection of upwardly propagating waves channelling energy from the chromosphere to the low corona}

\author{N. Freij$^{1}$, E. M. Scullion$^{2,3}$, C. J. Nelson$^{1,4}$, S. Mumford$^{1}$, S. Wedemeyer$^{2}$ and R. Erd\'elyi$^{1}$.}
\affil{$^{1}$Solar Physics \& Space Plasma Research Centre (SP$^{2}$RC), School of Mathematics and Statistics,\\ The University of Sheffield, Hicks Building, Hounsfield Road, Sheffield, S3 7RH U.K.}
\email{n.freij@sheffield.ac.uk}
\affil{$^{2}$Institute of Theoretical Astrophysics, University of Oslo, Postboks 1029 Blindern, 0315 Oslo, Norway.}
\affil{$^{3}$Astrophysics Research Group, School of Physics, SNIAM, Trinity College Dublin, Dublin 2, Ireland}
\affil{$^{4}$Armagh Observatory, College Hill, Armagh, BT61 9DG, Northern Ireland.}

\begin{abstract}
	There has been ubiquitous observations of wave-like motions in the solar atmosphere for decades. 
	Recent improvements to space- and ground-based observatories has allowed the focus to shift to smaller magnetic structures on the solar surface.
	In this paper, high-resolution ground-based data taken using the Swedish 1-m Solar Telescope (SST) \textcolor{red}{\textbf{is combined with co-spatial and co-temporal data from the}} Atmospheric Imaging Assembly (AIA) on-board the Solar Dynamics Observatory (SDO) satellite to analyse Running Penumbral Waves (RPWs).
	RPWs have always thought to be radial wave propagation that occur within sunspots.
	Recent research has suggested that they are in fact upwardly propagating field-aligned waves (UPWs).
	Here, RPWs within a solar pore are observed for the first time and are interpreted as UPWs due to the lack of a penumbra that is required to support RPWs.
	These UPWs are also observed co-spatially and co-temporally within several SDO/AIA elemental lines that sample the Transition Region (TR) and low corona.
	The observed UPWs are travelling at a horizontal velocity of around $17\pm0.5$ km s$^{-1}$ and a minimum vertical velocity of $42\pm21$ km s$^{-1}$.
	The estimated energy of the waves is around $150$ W m$^{-2}$, which is on the lower bounds required to heat the quiet-Sun corona.
	This is a new yet unconsidered source of wave energy within the solar chromosphere and low corona.
\end{abstract}


\section{Introduction}

	How energy is transported from the lower solar atmosphere into the corona is an important question that has yet to be fully answered despite decades of research\citep{erdelyi2004heating,erdelyi2007heating,taroyan2009heating}.
	The complex interactions between strong magnetic fields and powerful flows, the latter created by the interplay of gravity, convection and magnetic forces, leads to a number of dynamic phenomena throughout the atmosphere, such as magneto-hydrodynamic (MHD) waves \citep{Edwin1983}, which are theorised to supply energy into the corona.
	Strong inhomogeneities and steep gradients of key atmospheric properties (such as temperature and density) can lead to strong reflection of wave energy in the upper chromosphere.
	It has proved difficult to both observe \citep{Aschwanden2006,Marsh2006,Jess2009,Taroyan2009,McIntosh2011,Parnell2012,Morton2012,Wedemeyer2012,Mathioudakis2013} and simulate \citep{steiner1998,hasan2005,peter2006,erdelyi2007,ErdeyiFedun2010,Vigeesh2012} the propagation of energy from the lower atmosphere into the corona \citep{Vecchio2007,DePontieu2007,Zaqarashvili2009,DePontieu2011,mcintosh2012,Rutten2012}.
	
	The most basic model of MHD theory suggests that three distinct types of waves should manifest in the solar atmosphere; namely slow and fast magneto-acoustic and the widely sought-after Alfv\'en wave \citep{Banerjee2007,Jess2009,Suzuki2011,McLaughlin2011,McIntosh2011,Mathioudakis2013}.
	High spatial and temporal resolution observations carried out using modern ground- and space-based instrumentation have revealed a plethora of energetic, incompressible \citep{Aschwanden1999,DePontieu2007,Jess2009}, compressible \citep{Morton2012}, and significantly more complicated \citep{DePontieu2011,Wedemeyer2012}, oscillations and flows.
	What has yet to be observed is the direct propagation of energy from the lower regions of the solar atmosphere into the corona raising the question as to whether any of these wave processes are actually heating the outer solar atmosphere.
    Here, we contribute to addressing this question.
	
	Running penumbral waves (RPWs) were originally thought to be evidence of horizontal wave propagation\citep{Zirin1972,Giovanelli1972,Bloomfiel2008} which traced the topology of the local magnetic field \citep{Zhugzhda1973,Nye1974} around large sunspots.
	Due to this assertion, RPWs have been largely ignored with regards to any potential injection of energy into the corona.
	More recently, it has been suggested that these events are, in fact, upwardly propagating waves (UPWs)\citep{Bogdan2006,Bloomfiel2008,Jess2013}, which could facilitate the propagation of non-thermal energy into the corona.
	Here, we present the first observations of UPWs situated around a pore and demonstrate that these waves can indeed penetrate from the lower solar atmosphere into the corona, potentially making them an excellent candidate for plasma heating within solar active regions (ARs).
	
	We discuss here the propagation of UPWs through the plasma surrounding a large pore structure.
	By conducting a multi-wavelength, multi-instrument analysis, we are able to trace upward propagating wave-fronts from the chromosphere into the transition region (TR) and corona, estimating key properties such as apparent horizontal and vertical velocities, and non-thermal energy supply.
	The paper is organised as follows: Sect.\ref{sect1} details the collection and reduction of the data presented; Sect.\ref{sect2} describes the analysis of the data and studies the observed UPWs within the AR; Sect.\ref{sect3} we summarise and conclude.
	
\section{Data Collection and Reduction}
\label{sect1}

	\begin{figure*}
		\centering
		\includegraphics[scale=0.4]{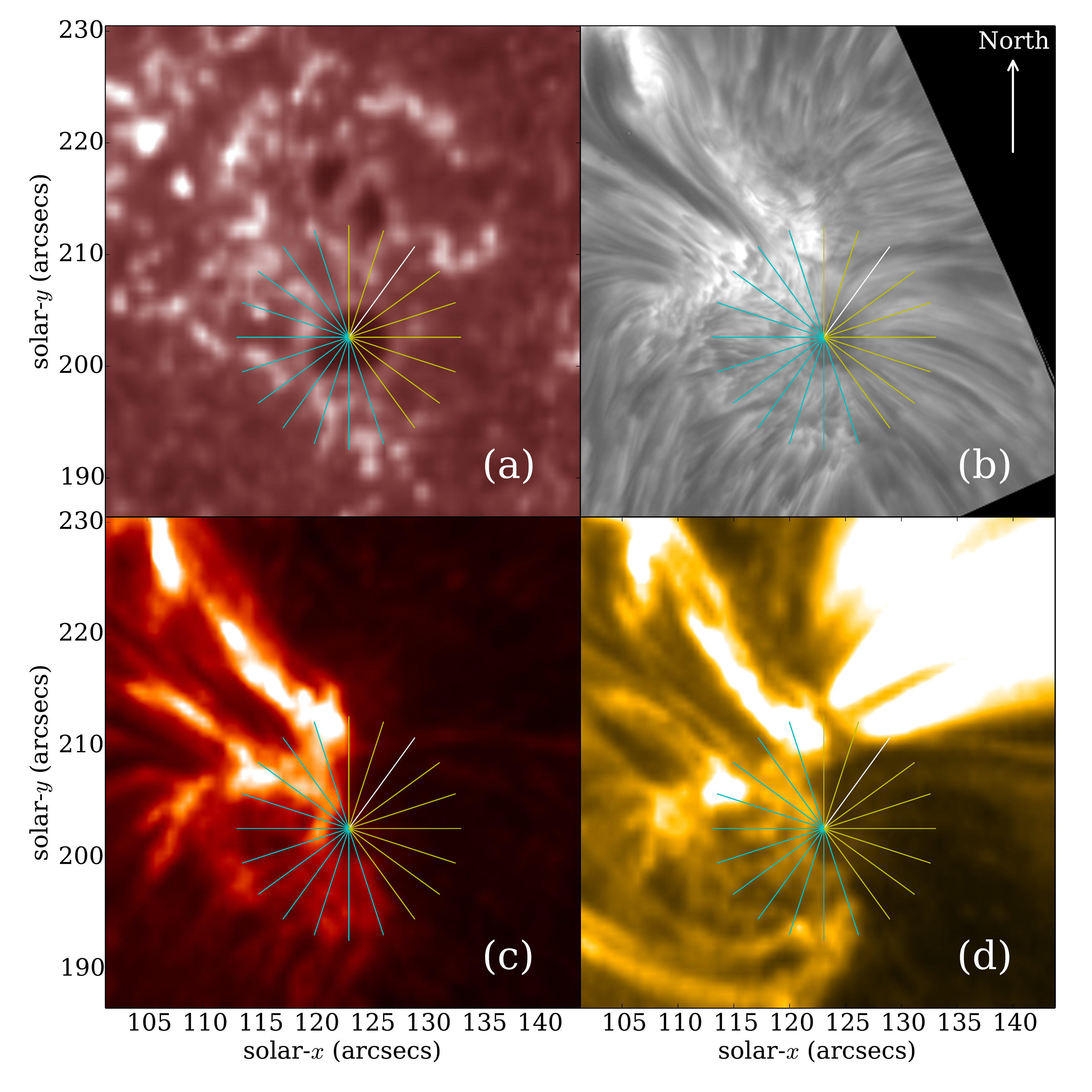}
		\caption
		{
		An overview of the field-of-view (FOV) inferred by SST/CRISP and SDO/AIA consisting of:
		(a) SDO/AIA $170$ nm, detailing the photosphere; (b) SST H$\alpha$ $656.28$ nm (line core) sampling the chromosphere; the  (c) SDO/AIA $30.4$ nm filter (TR); and the lower corona detailed by (d) SDO/AIA $17.1$ nm.
		The white line on each image represents the slit used to construct the time-distance diagrams plotted in Fig. \ref{fft_slit}.
		The yellow and cyan lines outline each slit used to investigate UPW behaviour.
		\textcolor{red}{\textbf{The yellow slits show where UPWs were observed and cyan slits show no UPWs.}}
		}
		\label{overview}
	\end{figure*}
	
	The analysis presented here is conducted on AR $11511$, which displayed a myriad of complex features during these observations.
	The ground-based data were obtained using the CRisp Imaging SpectroPolarimeter (CRISP)\citep{Scharmer2008} instrument, situated at the Swedish 1-m Solar Telescope (SST), on the 22nd June 2012 between $07:23$ UT and $08:28$ UT, during a period of excellent seeing.
	These data have a high spatial resolution of around $0.2$\arcsecs\ ($1$\arcsecs\ $\approx$ 725 km) and a cadence of $2.2$ seconds, allowing the small-scale structures of the lower solar atmosphere to be resolved (diffraction-limited) using a narrow-band $0.0269$ nm H$\alpha$ filter centred on $656.28$ nm.
	H$\alpha$ line scans were returned for $-0.1032, -0.0774, 0$ and $0.1032$ nm
	Each frame captured by the SST/CRISP instrument sampled a $68$\arcsecs\ by $68$\arcsecs\ FOV close to the disc centre.
	The data were reconstructed using the {\it Multi-Object Multi-Frame Blind Deconvolution} (MOMFBD) technique, giving an overall cadence of $2.2$ seconds and a spatial resolution of 0.12\arcsecs\citep{Noort2005}.
	\textcolor{red}{\textbf{We followed the standard procedures in the reduction pipeline for CRISP data (\citet{2014arXiv1406.0202D}) which includes the post-MOMFBD correction for differential stretching suggested by \citet{2012A&A...548A.114H}, also see \citet{2013ApJ...769...44S} for more details.}}
	
	Finally, co-aligned highly ionised plasma comprising the upper solar atmosphere was observed using the Solar Dynamics Observatory's (SDO) Atmospheric Imaging Assembly (AIA) instrument at a spatial resolution of approximately $1.5$\arcsecs\ and a temporal resolution of $12$ seconds. 

	In Fig. \ref{overview}, we include a general overview of the FOV analysed here, taken at $07:23$ UT.
	The pore of primary interest is located at \textcolor{red}{\textbf{approximately [123\arcsecs, 203\arcsecs]}} in helioprojective coordinates, and can be easily identified as it is situated underneath the overlaid cyan star symbol. 
	Four images sampled at different heights in the atmosphere are included to give an impression of the three-dimensional structuring evident in this region.
	The photosphere and chromosphere are sampled by the SDO/AIA $170$ nm filter (Fig. \ref{overview}a) and the SST/CRISP H$\alpha$ line core (Fig. \ref{overview}b), respectively.
	The dynamic fibril events which appear to protrude away from the large pore in the H$\alpha$ line core, obscure the majority of the large-scale structuring (such as the network) observed within the photosphere.
	Only in regions where strong vertical magnetic fields are present, such as within the confines of the large pore, does any evidence of the photospheric structuring penetrate into the chromosphere.
	Finally, the TR and corona are observed through the SDO/AIA $30.4$ nm (Fig. \ref{overview}c) and $17.1$ nm (Fig. \ref{overview}d) filters.
	It should be noted that two small pores are also within the FOV, situated at approximately \textcolor{red}{\textbf{[123\arcsecs, 215\arcsecs]}}, however, they are not evident in the H$\alpha$ line core.

\section{Results}
\label{sect2}

\subsection{The observed Active Region}
	\begin{figure*}
		\centering
		\includegraphics[scale=0.30]{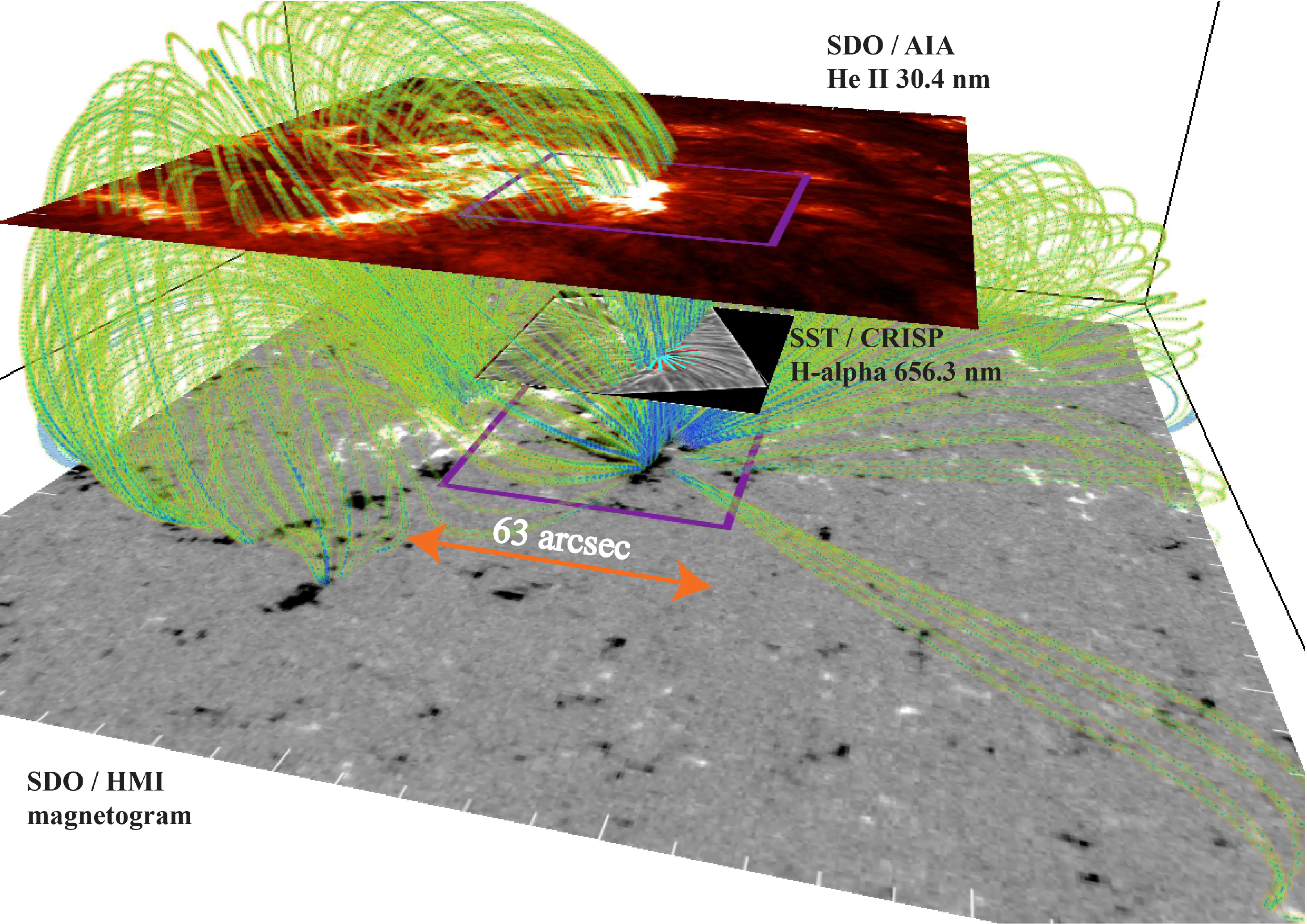}
		\caption
		{
		The base layer indicates the magnetic field inferred by the SDO/HMI instrument.
		The purple box highlights the SST/CRISP FOV which is overlaid.
		An extended FOV context image from the SDO/AIA $30.4$ nm filter is also included.
		The green lines are the visualisation of the magnetic field extrapolation.
		A strong correlation exists between these lines and the brighter regions in the SDO/AIA $30.4$ nm image underpinning that the extrapolation is a \textcolor{red}{\textbf{reasonable approximation over such a large height.}}
		}
		\label{mag_field}
	\end{figure*}

	In Fig. \ref{mag_field}, a stacked image outlining the coupling between the lower and upper regions of the solar atmosphere is presented.
	An extended FOV of the photospheric magnetic field is used as the base (with the SST/CRISP FOV overlaid as the purple box), from which the extrapolated field lines are plotted.
	Co-aligned photospheric magnetic field data were inferred by the SDO's Helioseismic and Magnetic Imager (HMI) instrument at a spatial resolution of around $1$\arcsecs\ and a cadence of $45$ seconds.
	Extrapolations of the magnetic field were then achieved by passing these data into the MPole Interactive Data Language package\citep{Longcope1996,Longcope2002}.
	
	We use MPOLE, to determine the 3D coronal magnetic field line connectivity about the FOV as observed by CRISP. MPOLE implements the Magnetic Charge Topology models and the Minimum Current Corona model  to derive the coronal field from a set of point charges. In our analysis, the charges are an approximation of an observed photospheric magnetic field. The complete set of charge positions and strengths (fluxes) are contained as a set poles. The poles are extracted from the observations through applying a feature tracking algorithm to HMI magnetograms of the active region of interest (extended about the CRISP co-aligned FOV by 50 arcsec in both solar-\textit{x} and solar-\textit{y} directions). Feature tracking of regions of positive and negative flux is carried out using YAFTA (Yet Another Feature Tracking Algorithm)\citep{DeForest2007}. Poles are labelled features which are collections of pixels in the magnetogram that are grouped according to criterion such as, spatial size and magnetic field strength. Subsequently, pixels below a threshold in flux density are not grouped, and receive a zero label in the mask. The thresholds are employed to ensure a suitably representative distribution of the  magnetic flux concentrations of the active region of interest.
	
	It is immediately noticeable that a non-rotationally symmetric distribution of field lines is present.
	Over-laid the magnetic field, we stack concurrent images from the SST/CRISP H$\alpha$, SDO/AIA $30.4$ nm, and SDO/AIA $17.1$ nm filters.
	Typically, the formation heights of the H$\alpha$ line core is estimated to be around $1.5$ Mm, which agrees to the mid-chromosphere \citep{Leenaarts2007}.
	The SDO/AIA $30.4$ nm and $17.1$ nm filters correspond to plasma in the TR and low corona, while SDO/AIA $19.3$ nm and $21.1$ nm filters correspond to plasma in the corona/hot flare plasma and AR corona, respectively.
	The chromosphere shows many elongated dark and bright structures surrounding the pore, identified as fibrils.
	Furthermore, a bright moss-like region to the north of the pore is evident, which corresponds well with regions of high magnetic flux, identified by the extrapolation process.
	The associated magnetic field from the large pore is observed to penetrate into the chromosphere and potentially higher, and corresponds well with the regions of increased intensity within the $30.4$ nm and $17.1$ nm filters, \textcolor{red}{\textbf{supporting that this extrapolation is reasonable over such a large height.}}
	\textcolor{red}{\textbf{The umbra of the two smaller pores do not appear to penetrate into the chromosphere, most likely due to insufficient magnetic flux.}}
	It should be noted, however, that UPWs patterns are still seen to propagate above the location of the rightmost pore in the H$\alpha$ line core.
	\textcolor{red}{\textbf{This indicates that the magnetic field lines do still expand into the solar chromosphere.}}
	In the higher temperature filters, the clarity of the pore fades, and large-scale loop structures, co-spatial with the extrapolated field lines, can be found.
	On the opposite side of the pore, a region of lower emission is observed in the TR and coronal lines co-spatially with less vertically inclined field lines returned by the magnetic field extrapolation.
	In the following sections, we discuss the influence of the magnetic field topology on observations of UPWs within this AR.
	It is imperative to note, that the height of each stacked image in Fig. \ref{mag_field} was estimated merely for ease of visualisation and should not, therefore, be used as strong evidence that the less vertically inclined field lines do not penetrate into the upper atmosphere. 

\subsection{Upwardly Propagating Waves}
	
	\begin{figure*}
		\centering
		\includegraphics[scale=0.4]{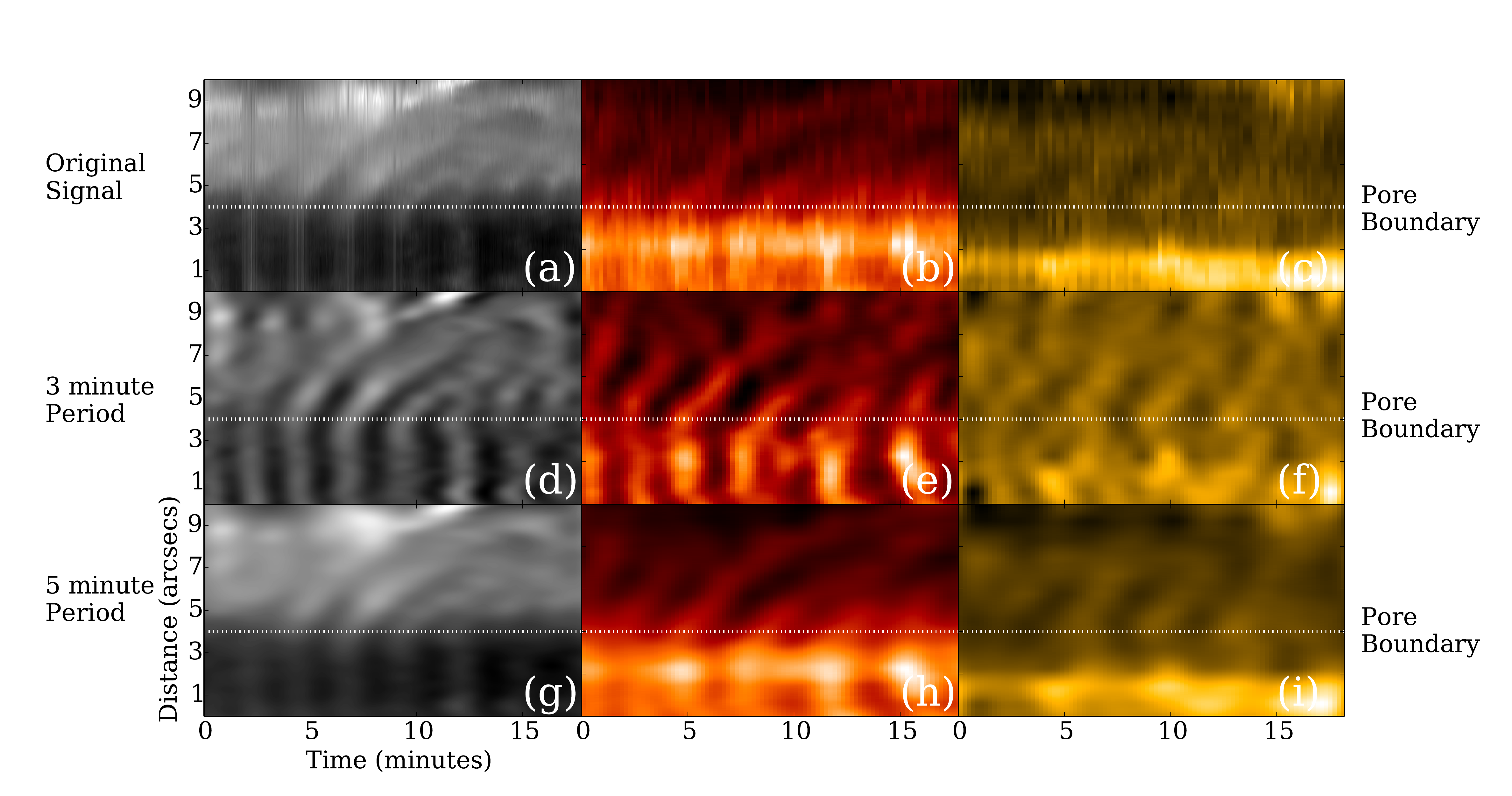}
		\caption
		{
		(Top row) Unfiltered time-distance slits for the H$\alpha$ line core (a), SDO/AIA $30.4$ nm filter (b), and $17.1$ nm filter (c) constructed for the white slit in Fig. \ref{overview}. (Middle row) Time-filtered 3-minute FFT output for H$\alpha$ (d), SDO/AIA $30.4$ nm (e), and SDO/AIA $17.1$ nm (f). (Bottom row) 5-minute FFT output for H$\alpha$ (g), SDO/AIA $30.4$ nm (h), and SDO/AIA $17.1$ nm (i).
		The windows used are centred on $3\pm1.5$ mhz (referred to as $5$ minutes) and $5\pm1.5$ mhz (referred to as $3$ minutes).
		\textcolor{red}{\textbf{The white dotted line is the pore boundary, below the line is the pore and above is the background chromosphere.}}
		}
		\label{fft_slit}
	\end{figure*}
	
	The main focus of this article is the analysis of UPWs.
	These events manifest as dark wavefronts, easily identified against the H$\alpha$ background, which appear to propagate radially away from the large pore with a coverage angle of approximately $160\degree$.
	The coverage of the UPWs is inclusive of both unstructured (such as at the north of the pore) and highly structured regions (on the east of the pore), implying that no specific magnetic topology is required in the H$\alpha$ line core to facilitate the propagation of these waves.
	It is interesting to note, however, that no UPWs are observed to propagate either south or west from the pore during these observations, implying that a fundamental, but as of yet unknown, factor is limiting either the observation or propagation of waves in this region.
	A reason for the absence could be the inclination of the magnetic field (see Fig. \ref{mag_field}) and will be expanded upon later in this Section. 
		
	In Fig. \ref{fft_slit}, we present a series of time-distance diagrams constructed using the white representative slit overlaid on Fig. \ref{overview}.
	The top row of Fig. \ref{fft_slit} plots the raw data extracted for this slit between 07:23:35 UT and 07:41:53 UT for the H$\alpha$ line core (a), the SDO/AIA $30.4$ nm filter (b), and the SDO/AIA $17.1$ nm filter (c).
	It should be noted that the start times for the SDO/AIA $30.4$ nm and $17.1$ nm filters are 9 seconds and 1 second ahead of the SST/CRISP data series, respectively.
	The UPWs are easily identified within the H$\alpha$ line core (as dark wavefronts) and the SDO/AIA $30.4$ nm filter (as bright wavefronts) propagating diagonally away from the pore between $3$\arcsecs\ and, approximately, $8$\arcsecs.
	The apparent horizontal velocity of the observed UPWs appears to decrease as the wavefront propagates away from the source.
	It has been hypothesised that the decrease in speed may be explained by ``the combined action of different frequency modes''\citep{Kobanov2006}, \textit{i.e.,} that an UPW is a superposition of two or more waves with different frequencies.
	Within the representative H$\alpha$ slit, the detected UPWs slow from $17\pm0.5$ km s$^{-1}$ to $12\pm0.5$ km s$^{-1}$ at distances of $4$\arcsecs\ to $5$\arcsecs, respectively.
	To conclusively test whether the observed deceleration was a physical property of the waves or a product of using straight slits for analysis, we conducted further research of time-distance diagrams constructed using curved slits, which traced fibril 	structures within the H$\alpha$ line core.
	Due to the occurrence of this deceleration in each analysed slit, we conclude that this behaviour of a reduction in apparent velocity is indeed a property of UPWs.
	Intuitively, as only two factors, namely the actual velocity and the angle of propagation, are required to formulate the apparent velocity, we are able to tentatively suggest that we observe either a physical slow-down of the wavefront or a change in the angle of propagation of these waves.
	
	The spatial occurrence of these waves is a further interesting point which requires discussion.
	Through the analysis of each cyan slit highlighted in Fig. \ref{overview}, investigation into how the behaviour of these waves changes spatially around the pore is feasible.
	At distances between $2$\arcsecs\ and $3$\arcsecs\ away from the pore boundary (indicated by the dashed white line in Fig. \ref{fft_slit}) for each individual slit, the apparent phase speed ranges from $10$-$20$ km s$^{-1}$ (\textit{i.e.}, approximately the sound speed in the chromosphere). 
	As UPWs are observed as single wavefronts, it is possible that the magnetic field topology is influencing the apparent horizontal velocity spatially around the pore.
	By overlaying the slits in which UPWs are observed onto the interpolated magnetic field, plotted in Fig. \ref{mag_field}, we are able to infer a spatial correlation between the apparently less vertically inclined magnetic fields and the occurrence of UPWs.
	The observations of such non-radially symmetric wavefronts around a pore, guided by the magnetic field, suggests that the extension of the magnetic field into the solar atmosphere from the pore, is non-axially symmetric.
	This result poses an interesting question: Does a combination of viewing angle and magnetic field topology limit the potential detection of propagating UPWs around the magnetic waveguide?
	It is imperative that a future analysis, ideally combining observations and simulations, be undertaken to further test this.  
	
    We now direct our investigation towards understanding the potential influence of different wave modes on the raw UPW signals.
	By employing the FFT technique on each row of the time-distance diagrams (Fig. \ref{fft_slit}a-c), the $3$-minute period for each wavelength can be isolated from the general wave behaviour.
	\textcolor{red}{\textbf{The windows used are Gaussian shaped, centred on $3\pm1.5$ mhz (referred to as $5$ minutes) and $5\pm1.5$ mhz (referred to as $3$ minutes) with a width of 2mHz.}}
	
	The second row of Fig. \ref{fft_slit} depicts the result of such an analysis for the H$\alpha$ line core (d), the SDO/AIA $30.4$ nm filter (e), and the SDO/AIA $17.1$ nm filter (f).
	The H$\alpha$ $3$-minute component starts off within the pore as an umbral flash-like event and, then, as the wave enters the surrounding atmosphere, moves away at a near constant speed, comparable to the raw data.
	It is easy to identify, that within the H$\alpha$ line core $3$-minute slit, the contrast of the waves against the background is increased when compared to the raw data.
	This suggests that the $3$-minute mode provides a high proportion of the energy carried by UPWs around the pore. 
	A similar behaviour is observed within the SDO/AIA $30.4$ nm wavelength, however, no signal is isolated within the SDO/AIA $17.1$ nm filter for this slit.
	Understanding these observations in terms of the physical properties of waves is essential to fully understand the UPW phenomena.
	Overall, the coverage angle, around the pore, of the $3$-minute mode within the SDO/AIA $17.1$ nm filter is approximately $50$ \% lower than the $30.4$ nm filter.
	The question as to whether this is a result of the waves not propagating into the $17.1$ nm passband or a reduced contrast against the background should lend itself to an interesting future study.

	\begin{figure*}
		\centering
		\includegraphics[scale=0.35]{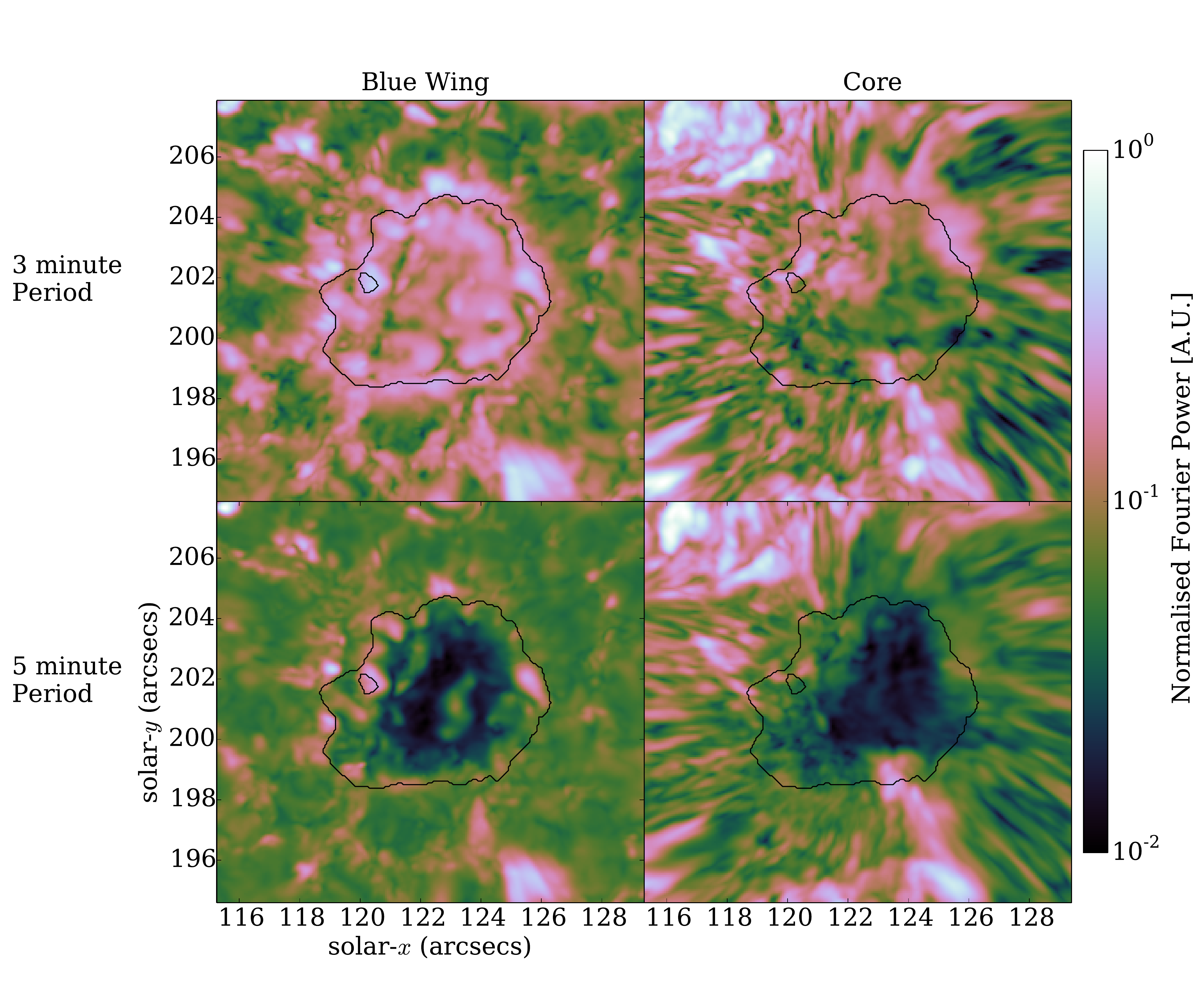}
		\caption{
		The spatial distribution of normalised Fourier power of the LOS intensity with 3- and 5-minute filter windows.
		The black contour line highlights the pore boundary as observed within the H$\alpha$ line wings. We depict the:
		(a) 3-minute filtering of the H$\alpha$ wing;
		(b) 3-minute filtering of the H$\alpha$ core;
		(c) 5-minute filtering of the H$\alpha$ wing;
		(d) 5-minute filtering of the H$\alpha$ core.
		}
		\label{fft_power}
	\end{figure*}

	Analysis of the $5$-minute period (Fig. \ref{fft_slit}g-i) allows for further inferences about the nature of these waves to be made.
	Within the H$\alpha$ line core, the occurrence of the $5$-minute mode is limited to regions outside of the pore, potentially due to the dependence of higher frequency modes on the magnetic field inclination\citep{DePontieu2004}.
	The phase speed is also reduced by approximately $1$-$2$ km s$^{-1}$ consistently around the pore.
	As there is more power within the $3$-minute mode close to the pore, it is assumed that this comprises the dominant component of the raw wavefront.
	It is possible, therefore, that the increased influence of the $5$-minute component as the wave moves away from the pore could explain the deceleration in raw phase speed, however, further research should be carried out to fully test this assertion.
	\textcolor{red}{\textbf{Within the SDO/AIA $17.1$ nm filter, the $5$-minute mode has a more defined wave pattern than the $3$-minute mode.}}
	We are, therefore, able to suggest that the $5$-minute mode more easily penetrates into the $17.1$ nm passband as has been suggested by previous researchers \citep{Moortel2002}, potentially providing energy into the TR.

	Another method that can be exploited to further understand the physical properties of these waves is a time-delay analysis.
	We were able to compare both the raw and FFT-filtered data for each wavelength in order to establish whether evidence of a lag exists.
	By taking into account the different start times for the SST/CRISP and SDO/AIA data, no observable lag was discernible.
	Therefore, we are able to conclude that either any lag between the signals is less than the cadence of these SDO/AIA data or that, indeed, no lag exists.
	Should the second hypothesis prove true, it would suggest that these observations support the propagation of a single wave, which occurs within the combined passbands of each of these filters, {\it i.e.}, around the TR.
	
	By expanding the FFT analysis to the full FOV, we are able to analyse how power is manifested within the local plasma.
	Fig. \ref{fft_power} shows the result of applying a $3$- and $5$-minute period FFT filter on the LOS intensity for the H$\alpha$ line core and far wing ($-0.1032$ nm).
	The same process was also applied to the concurrently taken SDO/AIA data, however, the obtained power maps lost their spatial structure and, as such, we were unable to make further conclusions.
	The black contour depicts the outline of the pore as observed in the photosphere sampled by the H$\alpha$ wing.
	Within the photosphere (Fig. \ref{fft_power}a,c) the 3-minute power is isolated inside the pore structure; specifically, there appears to be large regions of power tracing the boundary of the pore, apparently analogous to the distribution of power within a sunspot \citep{Stangalini2012,Reznikova2012}.
	The power in the 5-minute band is minimal in the body of the pore but their is an increase at the pore-photosphere transition boundary corresponding to enhanced \textit{p}-mode power \citep{Mathew2008}.
    We interpret the confinement of the power within the pore as evidence that UPWs are driven by $p$-modes propagating vertically within the pore, which acts as a magnetic waveguide.
	
	Finally, we are able to analyse the H$\alpha$ line core.
	The increase of power especially within the 3-minute, easily observed to the north-east of the pore, corresponds well with the occurrence of UPWs within these data.
	It is intuitive to suggest that, as the FFT analysis is only applied in the vertical direction, the horizontal component of the UPWs in these regions limits the detection of power.
	Potentially, the increase in the FFT power observed to the north of the pore, could be indicative of the propagation of UPWs into the upper solar atmosphere along more vertically inclined magnetic field lines (as observed within Fig. \ref{mag_field}).
	We interpret the lack of power co-spatially with the UPWs (in the east) as further evidence that the pore's magnetic field has become non-symmetric in the chromosphere.
	Evidence of the apparent dependence of both the observation of UPWs and the localised power within the plasma around a pore on the potential magnetic field topology, as presented within this article, is a key step in fully understanding the complex nature of coupling between layers of the solar atmosphere.

\subsection{Energy of UPWs}

	Following the identification and detailed analysis of UPWs around a pore, it is essential to estimate the potential energy carried by these waves into the upper solar atmosphere.
	Due to the decrease and increase in intensity in comparison to the background plasma for the H$\alpha$ line core and the SDO/AIA filters, respectively, it can be inferred that the wavefront represents an increase in density \citep{Allen1947,Leenaarts2012}.
	By measuring the contrast between the background plasma and the wavefronts, it is apparent that the intensity perturbations are within the linear regime and, therefore, these waves appear to be magneto-acoustic in nature.
	In order to further this analysis, we assume here that the lack of observed time-delay in these data implies that the lag is below the cadence of these data.
	Given estimated formation height-differences between the chromospheric H$\alpha$ line core and the SDO/AIA $30.4$ nm filter can be estimated to be around $0.5\pm0.25$ Mm, the upward propagation speed can be calculated as $42\pm21$ km s$^{-1}$.
	This speed is close to previous estimates of the fast speed in the chromosphere \citep{Morton2012}.  
	It should be noted, that this corresponds well with previous results, which suggest that $p$-mode oscillations, which appear to drive these UPWs, are converted to fast modes \citep{Vigeesh2012}.
	The combination of these factors allows us to suggest that one of the most likely interpretations of these observations is that UPWs are {\it fast sausage} waves.
				
	With the wave type being identified, it is now possible to calculate the estimated non-thermal energy for these waves.
	It is possible to estimate the energy flux at each pixel based on linearised MHD theory \citep[e.g.][]{Kitagawa2010}.
	The equation for the total energy flux of the fast MHD sausage wave is
	\begin{equation}
		E_{wave} = \sum\limits_{i = 1}^{N}\rho_{0}[\tilde{I_{i}}/I_{0}]^{2}c_{ph}^{3},
		\label{energy}
	\end{equation}
	where $\tilde{I_{i}}$ is the intensity perturbation for each pixel, $I_{0}$ is the background intensity, $c_{ph}$ is the phase speed of the sausage wave, $\rho_{0}$ is the background density.
	We sum over each pixel which is part of the wave, giving us the average energy for that wave.
	Since the wave is a fast MHD sausage wave, the phase speed is $c_{fast}$ which is the local fast speed, however, since the ratio of the Alfv{\'e}n to the sound speed is $>>1$, the Alfv{\'e}n speed is the dominant value in the fast speed calculation.
	This assumes that the plasma is optically thin (intensity is proportional to density), which is true for the coronal lines however, not the case for H$\alpha$.
		
	This analysis leads to energy estimates of the order of $150$ W m$^{-2}$ for the wavefronts in the H$\alpha$ line core.
	These values drop by two orders of magnitude within the SDO/AIA filters. 
	These energy flux values are about a factor of $100$ less than reported for other abundant sausage wave events in the chromosphere\citep{Morton2012}, however, they still comprise a important fraction of the energy flux required to heat the local quiet \citep{Wedemeyer2012} and active corona \citep{Aschwanden2007}, respectively. 
	It should be noted, that these estimates are influenced by a number of observational factors, such as attenuation in the telescopic apparatus, changes in light levels throughout these data, and the angle of observation, to name a few.
	We do, however, suggest that during the period of these observations, there are approximately constant seeing conditions and, therefore, these energy estimates should be consistent.
	\textcolor{red}{\textbf{Magnetic pores cannot heat the entire corona, but can contribute to heating the local corona that is above and near the pore.
	The value for the energy flux is for the region where we can observe the UPWs and the most logical case is that UPWs occurs across the entire pore but are difficult to observe due to the local solar atmosphere.
	This should raise the value for the energy flux that has been obtained.}}

\section{Discussion}
\label{sect3}

	The results presented in this article support the assertions that waves propagating radially away from concentrated magnetic waveguides (such as pores and sunspots) in the solar photosphere have significant vertical components that give rise to the illusion of horizontal propagation. 
	The magnetic field reconstruction (as seen in Fig. \ref{mag_field}) gives us a useful insight into the non-radially symmetric nature of this pore and, specifically, how the apparent topology of the magnetic field influences UPWs.	
	The case that RPWs are in fact UPWs that travel along the field lines is mounting \citep{Bloomfiel2008,Jess2013}.
	Here, strong evidence is presented that energy from $p$-modes in the lower solar atmosphere travels directly upwards into the TR and lower corona.
	It has been reported that there is absorption of power at the boundary of the umbra-penumbra for a sunspot \citep[e.g.][]{Gosain2011}.
	Here, we observe enhanced power at the boundary of the pore at both three and  five minutes, while in the chromosphere, where UPWs are observed, there is a reduction of power.
	As the energy from the acoustic $p$-modes is converted into MHD waves along the flux tube, the period of the $p$-mode becomes three minutes and traces the magnetic field.
	When the wave travels into the TR and solar corona, there is decrease of the wave period.
	Rudimentary energy flux calculations reveal that these waves are able to contribute to heating the local corona, however, how much they contribute requires further study. 

	From this primarily wave-based study of the solar atmosphere we deduce that, in the outside environment surrounding the pore, the magnetic field of the pore becomes non-symmetric.
	The non-symmetric magnetic field appears to be integral in allowing UPWs to be observed, however, whether these events occur in other regions around the pore but are undetected, requires further study.
	Further investigation is also required to fully assess whether the lack of UPW signal within some regions around the pore is a consequence of seeing or an, as of yet unascertained, physical property (such as the cut-off frequency). 
	\textcolor{red}{\textbf{A possible interpretation of these waves is a singular wavefront observed in multiple pass bands, data from a wider range of sources should help answer these.}}
	This calls for an extensive investigation using detailed spectropolarimetry (ground-based) data to resolve the issue but also to determine the consequence of changing the LOS (i.e on the limb) on the observation of UPWs. 
	We have shown that the complex lower solar atmosphere, which does act as a powerhouse in the heating of the outer atmosphere, can in fact be further understood through a purely wave-based investigation.
	
\section*{Acknowledgements}
	This work is supported by the UK Science and Technology Facilities Council (STFC).
	Research at the Armagh Observatory is grant-aided by the N. Ireland Dept. of Culture, Arts nand Leisure.
	This research has made use of the Solar Dynamics Observatory, joint ESA and NASA funded JHelioviewer, NASA’s Astrophysics Data System, and the Virtual Solar Observatory.
	The Swedish 1-m Solar Telescope is operated on the island of La Palma by the Institute for Solar Physics of Stockholm University in the Spanish Observatorio del Roque de los Muchachos of the Instituto de Astrof\'{\i}sica de Canarias.
	We thank Luc Rouppe van der Voort (Institute of Theoretical Astrophysics, University of Oslo) and J. de la Cruz Rodriguez (University of Uppsala, Sweden) for data reductions with MOMFBD for SST/CRISP and standard SolarSoft routines for SDO/AIA.
	The authors would like to acknowledge the SunPy, NumPy, SciPy, Matplotlib and scikit-image, python projects for providing computational tools to analyse the data.
	RE acknowledges M. K\'eray for patient encouragement and is also grateful to NSF, Hungary (OTKA, Ref. No. K83133).
	
\bibliographystyle{apj}
\bibliography{UPWs}

\end{document}